\documentclass[prb,
twocolumn,
superscriptaddress,showpacs,amsmath,amssymb]{revtex4}
\usepackage{amsfonts}
\usepackage{bm}
\usepackage{verbatim}
\usepackage{color}
\usepackage{graphicx}

\newcommand{\im}{i}
\newcommand{\unitvec}[1]{\hat{\mathbf{#1}}}
\newcommand{\vek}[1]{\mathbf{#1}}

\begin{document}

\title{Effective Minkowski to Euclidean signature change of the magnon BEC pseudo-Goldstone mode in polar $^3$He}
\author{J. Nissinen}

\affiliation{Low Temperature Laboratory, Aalto University, P.O. Box 15100, FI-00076
Aalto, Finland}

\author{G.E.~Volovik}

\affiliation{Low Temperature Laboratory, Aalto University, P.O. Box 15100, FI-00076
Aalto, Finland}

\affiliation{Landau Institute for Theoretical Physics, acad. Semyonov av., 1a,
142432, Chernogolovka, Russia}

\date{\today}

\begin{abstract}
We discuss the effective metric experienced by the Nambu-Goldstone mode propagating in the broken symmetry spin-superfluid state of coherent precession of magnetization. This collective mode represents the phonon in the RF driven or pulsed out-of-equilibrium Bose-Einstein condensate (BEC) of optical magnons. We derive the effective BEC free energy and consider the phonon spectrum when the spin superfluid BEC is formed in the anisotropic polar phase of superfluid $^3$He, experimentally observed in uniaxial aerogel $^3$He-samples. The coherent precession of magnetization experiences an instability at a critical value of the tilting angle of external magnetic field with respect to the anisotropy axis. From the action of quadratic deviations around equilibrium, this instability is interpreted as a Minkowski-to-Euclidean signature change of the effective phonon metric. We also note the similarity between the magnon BEC in the unstable region and an effective vacuum scalar ``ghost" condensate.
\end{abstract}
\pacs{
}

\maketitle

%\section{Introduction}

\section{Introduction}

There are different classes of broken symmetry states that experience the phenomenon of spin superfluidity \cite{Sonin2010}. The first of them contains magnetic systems with spontaneously broken continuous symmetry of spin rotations, $SO_S(3)$ or the planar subgroup $SO_S(2)$. The spontaneous breaking of this symmetry leads to the associated Nambu-Goldstone (NG) modes  (spin waves or magnons), to spin supercurrents and to topological defects, such as spin vortices with spin supercurrent circulating around the cores. Examples are provided by some solid state magnetic materials \cite{AndreevMarchenko1980,Brataas2017}, and by spin-triplet superfluid phases of liquid $^3$He \cite{VollhardtWolfle1990,Volovik2003}. In superfluid $^3$He the spin-orbit (dipole) interaction is tiny, and the spin rotation $SO_S(3)$ symmetry is almost exact. The broken $SO_S(3)\times U(1)$ symmetry of the superfluid order parameter leads to the recently observed half-quantum vortices \cite{Autti2016a}, which have  both  spin and mass supercurrent circulation around the vortex cores.

Due to spin-orbit interaction, which explicitly violates the spin rotation symmetry, some magnons acquire small masses and become pseudo NG modes.
In high energy physics, the formation of such a massive boson is called the Little Higgs scenario 
\cite{LittleHiggsReview2005}, which
may explain why the Higgs boson has relatively small mass of 125 GeV.  In $^3$He-B the parametric decay of 
optical magnons to pairs of light Higgs modes has been observed \cite{Autti2016b}. In this scenario, the spin and spin-mass vortices become the termination lines of topological spin solitons \cite{MineevVolovik1978}. In the polar phase of $^3$He, the topological soliton emerges between two neighboring half-quantum vortices when the magnetic field is tilted with respect to an anisotropy axis and is resolved in NMR experiments \cite{Autti2016a}. 

The second class of spin superfluid states encompasses states which are periodic in time. A state with  spontaneously formed phase-coherent precession of magnetization has been first observed 
in  $^3$He-B \cite{HPDexp,HPDtheory}. The lifetime of this coherent precession  is extremely large  compared with thermalization time, and if dissipation is neglected, this spontaneously time-periodic state represents an example of a time crystal \cite{Wilczek2013,Volovik2013,TimeCrystalsReview2017}. From a different point of view, the spontaneously formed coherent precession can be considered in the language of an out-of-equilibrium Bose-Einstein condensate (BEC) of quasiparticles \cite{BunkovVolovik2013}, which for the case of $^3$He-B are optical magnons.
The spontaneous breaking of time translation symmetry leads to spin current Josephson effect, to quantized vortices in the magnon BEC and to the new NG mode -- the propagating oscillations of the phase of precession \cite{Fomin1978, Fomin1980, Fomin1986,Fomin1990,Bunkov1986,Dmitriev2005,Skyba2008,Skyba2012}, which represents the usual phonon mode of the magnon BEC in the out-of-equilibrium BEC language \cite{BunkovVolovik2013}.

In experiments, the out-of-equilibrium magnon BEC in superfluid $^3$He arises when the system is either continuosly driven with an external transverse RF magnetic field $\mathbf{H}_{\rm rf} 
\perp \vek{H}$  or after a short transverse RF field pulse is applied. In the pulsed NMR experiment, after the RF pulse is turned off, the spin precession experiences dephasing due to inhomogeneity of the underlying superfluid texture. But then the phase coherence is rapidly restored  due to spin supercurrents, and the spins enter a long-lived state  -- the magnon BEC, where the macroscopic  
spin ${\bf S}$ is freely precessing at an angle $\beta$ with respect to $\vek{H}$, with the off-diagonal order parameter 
\begin{align}
\langle \hat{S}^{+} \rangle = \langle \hat{S}^x + \im \hat{S}^y \rangle = S_{\perp} e^{\im \omega t + \im \alpha}, \\
S_{\perp} = S \sin \beta, \quad n = \frac{S - S_{z}}{\hbar}.
\label{eq:precessing magnetization}
\end{align}
Here $n =S(1-\cos \beta)/\hbar$ is the magnon number density in the condensate, $\alpha$ the condensate phase, and the precession frequency $\omega$ plays the role of a chemical potential: 
$\mu \equiv \omega= \omega_{\rm rf}$ in the presence of continuous RF pumping (in thermodynamics this is the regime of the fixed chemical potential), and $\mu \equiv \omega$ in pulsed RF fields, where the global precession frequency $\omega$  is determined by the number $N$ of magnons pumped during the pulse (the regime of the fixed number of magnons). 

The nonequilibrium superfluidity of magnon BEC has also been observed in Yttrium Iron Garnet films \cite{Demokritov2008,Bozhko2016,Pokrovsky2016}   

In this paper we study the NG mode of the magnon BEC in the polar phase. The rest of this paper is organized as follows. In Sec.\ref{sec:Magnon specrum} we review the magnon spectrum in polar $^3$He and discuss the precessing magnon BEC and its phonon spectrum in Sec.\ref{sec:Dynamics}. In Sec.\ref{Sec:acoustic}, we analyze the acoustic phonon metric and identify the Minkowski-to-Euclidean signature change. In Sec\ref{Sec:HamiltonianToAction}, we compute the effective metric of quadratic deviations around equilibrium and conclude with an outlook in Sec.\ref{sec:Outlook}.

\section{Magnon spectrum and effective metric in the polar phase}
\label{sec:Magnon specrum}

The polar phase can be stabilized by immersing superfluid $^3$He in an uniaxially anisotropic aerogel, where the orbital anisotropy $\unitvec{n}$ of the condensate aligns along the aerogel strands. The order parameter of the polar phase is given as \cite{VollhardtWolfle1990}
 \begin{equation}
 A_{\alpha i}= \Delta_P \hat d_\alpha  \hat n_i e^{\im \Phi}\,.
\label{Polar}
\end{equation}
Here $\Delta_P$ is the gap amplitude with phase $\Phi$; $\hat{\mathbf{n}}$ the fixed orbital anisotropy along the aerogel strands; $\hat{\vek{d}}$ the unit vector of the spin-anisotropy axis of the Cooper pairs.
 The polar phase represents the superfluid analog of spin-nematic state in antiferromagnets \cite{AndreevMarchenko1980}, since the states $\hat{\vek{d}}$ and $-\hat{\vek{d}}$ can be connected by the change of the phase $\Phi$ by $\pi$. The latter gives rise to the half-quantum vortices, which have been observed in the polar phase \cite{Autti2016a}. 

Spin dynamics is governed by the Leggett equations for ${\bf S}$ and $\hat{\bf d}$, i.e. the `adiabatic' Hamiltonian $F$, which is the superfluid $^3$He free energy in the London limit:
\begin{align}
F &= F_{\rm spin} + F_{\rm grad} + F_{\rm so}, \\
f_{\rm spin} &= \frac{1}{2}\gamma_m^2 \mathbf{S} \boldsymbol{\chi}^{-1} \mathbf{S} - \gamma_m \mathbf{H}\cdot \mathbf{S} \\
f_{\rm grad} &=  
\frac{1}{2} K_{ij} \nabla_i \hat{d}_{\alpha} \nabla_{j}\hat{d}_{\alpha}
\label{Gradients}
\\
f_{\rm so} &= 
g_D(\unitvec{d}\cdot \unitvec{n})^2. 
\label{SpinOrbit}
\end{align}
In the polar phase, the spin susceptibility is given as $\chi_{\alpha\beta} = \chi_{\parallel} \hat{d}_\alpha\hat{d}_\beta+ \chi(\delta_{\alpha\beta} -\hat{d}_\alpha\hat{d}_\beta)$ and the spin orbit and gradient energy for the spin-vector $\unitvec{d}$ have  $K_{ij} = K_{\parallel} \hat{n}_i\hat{n}_j + K_{\perp}(\delta_{ij}-\hat{n}_i\hat{n}_j)$ and  $g_D=\frac{\chi \Omega_P^2}{2\gamma_m^2}$, where $\Omega_P$ is the Leggett frequency of the polar phase ($\gamma_m$ is the $^3$He nuclei gyromagnetic ratio). It follows that $\vek{S}= \chi \vek{H}/\gamma_m$ in equilibrium.

When the spin-orbit interaction $f_{\rm so}$ is neglected, the spectrum of  longitudinal and optical magnon modes, with polarizations $a=0,+1$ respectively, can be written in relativistic form:
\begin{equation}
g_S^{\mu\nu}p_\mu p_\nu + M_a^2  =0\,.
\label{MagnonSpectrum}
\end{equation}
Here $p_\mu$ is the 4-momentum of magnons, $p_\mu= (\omega, k_i)$ and $g_S^{\mu\nu}$ is the effective magnon metric -- the magnonic counterpart of the acoustic metric \cite{Unruh1981}.
In the polar phase, for the homogeneous static superfluid state, it takes the form
\begin{align}
g_S^{mn}=  c^2_{\parallel S} \hat n^m\hat n^n 
+ c^2_{\perp S} (\delta^{mn}-\hat n^m\hat n^n ) \,\,, \,\,  g_S^{00}=-1\,,
\label{EffectiveMagnonMetric}
\end{align}
where  the ``speeds of light" for magnons propagating parallel and transverse to $\hat{\mathbf{n}}$, respectively (this anisotropy has been measured \cite{Zavjalov2015})  are:
\begin{equation}
c^2_{\parallel S}=\gamma_m^2 K_{\parallel}/\chi \,\,,\,\, c^2_{\perp S}=\gamma_m^2 K_{\perp}/\chi  \,.
\label{SU2gauge}
\end{equation}
In Eq.(\ref{MagnonSpectrum}) the magnetic field $\vek{H}$ is chosen along $\unitvec{z}$, and the``invariant masses" are respectively:
\begin{equation}
M_+ = \omega_L  \equiv \gamma_m H \,\,,\,\, M_0=0\,,
\label{OpticalMagnonMass}
\end{equation}
where $\omega_L = \gamma_m H$ is the Larmor frequency. 

With the spin-orbit interaction (\ref{SpinOrbit}) the spectrum of magnons:
\begin{eqnarray}
\omega_0^2 = g_S^{mn}k_m k_n, \\
\omega_+^2 =  \Omega_{\rm P}^2 +\omega_L^2+ g_S^{mn}k_m k_n 
\,.
\end{eqnarray}

\section{Dynamics of the coherently precessing state}
\label{sec:Dynamics}

\subsection{Magnon BEC and phonon Hamiltonian.}

We are interested in the low-frequency and long wavelength dynamics of magnon BEC, which is developed in the background of the fast precession. This dynamics is described by the slow variables, magnon density $n$ of the condensate in Eq.(\ref{eq:precessing magnetization}) and the phase 
$\alpha$ of precession. As in conventional BECs, these two variables are canonically conjugated, and the linearized equations for these variables describe the Goldstone mode of the coherent precession -- the phonon propagating in the magnon condensate. 

The  Hamiltonian $H(n,\alpha)=H_{\rm BEC}-\mu N = F_{\rm BEC}$ for the slow magnon BEC modes can be obtained by averaging the spin-orbit and gradient terms, Eqs. \eqref{SpinOrbit} and \eqref{Gradients}, over the fast Larmor precession. We assume here that $\Omega_{\rm P}^2\ll \omega_L^2$, then to the zeroth order approximation, one has the pure Larmor precession at $\omega= \omega_L$, which can be expressed in the general form:
\begin{align}
\vek{S}(t) &= \vek{O}^{-1}(t) \vek{R} \vek{O}(t) S\unitvec{z}\,,  \hspace*{0.2\textwidth}
\label{Svector}\\ 
\hat{\bf d}(t) &={\bf O}^{-1}(t) {\bf R} {\bf O}(t) \hat{\bf x}\,, \quad  {\bf S}(t) \cdot  \hat{\bf d}(t)=0\,.
\label{dvector}
\end{align} 
Here ${\bf O}(t) =R_z(\omega_L t)$ is the transformation to the frame rotating with the Larmor frequency $\omega_L=\gamma H$,
and $\vek{R} = R_z(\alpha)R_y(\beta)R_z(\gamma)$ is the matrix of spin rotation in that frame
with  Euler angles $\alpha, \beta, \gamma$.
Abbreviating $s(x)\equiv \sin x$ and $c(x)\equiv \cos x$, one obtains explicit time dependence of $\vek{S}(t)$ and 
$\hat{\bf d}(t)$ in  the Larmor precession 
\begin{align}
\textstyle
\unitvec{S}(t) 
& = \textstyle c(\alpha-\omega_L t) s(\beta) \hat{\vek{x}} + s(\alpha - \omega_L t) s(\beta) \hat{\vek{y}} + c(\beta) \hat{\vek{z}}  ,
\nonumber
%\label{Svector2}
\\
\textstyle
\hat{\vek{d}}(t) 
%\nonumber \\
&= \textstyle [c(\beta) c(\alpha -\omega_L t ) c(\gamma + \omega_L t ) \!-\! s(\alpha -\omega_L t ) s(\gamma +\omega_L t )]\hat{\vek{x}} \nonumber\\
\textstyle
& \textstyle +[c(\beta) s(\alpha -\omega_L t) s(\gamma +\omega_L t)\!+\!c(\alpha -\omega_Lt ) s(\gamma +\omega_L t)]\hat{\vek{y}}  
\nonumber
\\
\textstyle
&\textstyle -s(\beta) c(\gamma +\omega_L t)\hat{\vek{z}}.
\nonumber
%\label{dvector2}
\end{align}

Averaging of the spin-orbit term   (\ref{SpinOrbit})  over the fast precession gives
%\begin{widetext}
\begin{align}
 \langle f_{\rm so}(t) \rangle =  \frac{g_D}{4} \bigg( 1+ \cos^2\lambda + (1- 3 \cos^2\lambda) \cos^2\beta - \label{SOaverage} \\ 
\frac{1}{2} (1+\cos\beta)^2 \sin^2\lambda \cos(2(\alpha+\gamma)) \bigg), \nonumber
\end{align}
where $\lambda$ is  the angle of the vector of  orbital anisotropy with respect to the static magnetic field,
 $\hat{\bf n}= \hat{\bf y} \sin \lambda +  \hat{\bf z} \cos \lambda$. 
 Minimization over $\alpha+\gamma$ gives  $\alpha+\gamma=0$ and, as a result, one obtains the following nonlinear contribution to the energy density of the condensate in terms of magnon density $n=S(1-\cos\beta)$:
\begin{align}
 \epsilon(n) &=  \langle f_{\rm so}(t) \rangle_{\gamma=-\alpha} \nonumber \\
 &= \frac{g_D}{4} \bigg( 1+ \cos^2\lambda + (1- 3 \cos^2\lambda) \cos^2\beta \label{SOaverageMin}\\
&\phantom{=} - \frac{1}{2} (1+\cos\beta)^2 \sin^2\lambda \bigg) \nonumber.
\end{align}
%\end{widetext}
The average over the gradient term follows similarly. Taking into account that $\gamma=-\alpha$ at equilibrium, one obtains
\begin{align}
\langle \nabla_i \hat{\bf d}(t)  \cdot \nabla_j \hat{\bf d(t)}  \rangle \hspace*{0.2\textwidth} \nonumber
\\
= \frac{1}{2} (1- \cos\beta)(3-\cos\beta)\nabla_i \alpha\nabla_j \alpha + \frac{1}{2} \nabla_i \beta \nabla_j \beta\, .
\label{Gradient2}
\end{align}
Finally, gathering all terms, we arrive to
%\begin{widetext}
\begin{align}
H(\alpha, n) = \int d^3 \vek{r}~ \frac{1}{4S^2}n(n_{\rm max}+n)K_{ij} \nabla_i \alpha \nabla_j \alpha 
\nonumber
\\
+ \frac{K_{ij}}{4n(n_{\rm max}-n)}\nabla_i n \nabla_j n 
\nonumber
\\
+ (\omega_L -\mu)n+  \epsilon(n)  + \gamma_m H_{\rm rf} S \sin \beta \frac{\alpha^2}{2}
\label{eq:Hamiltonian0}
\end{align}
%\end{widetext}
Here $n_{\rm max}=2S$.
We also added the  symmetry breaking term for small $\alpha$, which appears in case of cw-NMR and comes from the driving RF field $\vek{H}_{\rm rf} \parallel \hat{\vek{x}}$,
\begin{align}
f_{\rm sb}(\alpha, \beta) = -\gamma_m \vek{H}_{\rm rf}\cdot \vek{S} = -\gamma_m H_{\rm rf}S\sin\beta \cos \alpha \,.
\end{align}
It gives the mass to the phonon propagating in magnon BEC, see Eq.(\ref{PhononMass}). For small $n \ll n_{\rm max}$, the phonon Hamiltonian Eq. \eqref{eq:Hamiltonian0} transforms to the Ginzburg-Landau free energy $F_{\rm BEC}$ of the magnon BEC in the polar phase (see Eq.(\ref{GLgradient}) in the Appendix), where the precession averaged spin-orbit interaction $\epsilon(n, \lambda)$ serves as the interaction between the magnons in the BEC.

\subsection{Goldstone mode spectrum.}

Introducing the dimensionless variable $\tilde n=1-\cos\beta$, the Poisson brackets $\{\tilde n({\bf r}_1), \alpha ({\bf r}_2)\}=S^{-1}\delta({\bf r}_1 - {\bf r}_2)$ give the following equations of motion
\begin{equation}
\dot{\tilde n} = -\frac{1}{S}\frac{\delta H}{\delta \alpha} \,\,, \,\,  \dot\alpha= \frac{1}{S}\frac{\delta H}{\delta \tilde n},
\label{MotionEquations}
\end{equation}
from which in linear order in $\alpha$ and $\delta n$ one obtains the phonon wave equation
\begin{eqnarray}
\frac{\partial^2\alpha}{\partial t^2}= \epsilon''
\left[\gamma_{ij}\nabla_i \nabla_j\alpha
- \gamma_m H_{\rm rf}S\sin \beta\, \alpha  \right]
\label{WaveEquation1}
\\
- \frac{1}{4S^2} \frac{2+\tilde n}{2-\tilde n} K_{mn}K_{ij}\nabla_i \nabla_j\nabla_m \nabla_n\alpha \,,
\label{WaveEquation2}
\\
\epsilon''=\frac{d^2\epsilon}{dn^2}=\frac{1}{S^2} \frac{g_D}{4}   (1- 5 \cos^2\lambda) \,, 
\label{epsilon''}
\\
\gamma_{ij}=\frac{1}{2} (1- \cos\beta)(3-\cos\beta)K_{ij}
\label{gamma}
\end{eqnarray}

 Let us first neglect the 4th order term in Eq. \eqref{WaveEquation2}, then using  Eq. \eqref{WaveEquation1} one obtains the ``relativistic" spectrum of the NG mode -- the phonon propagating in magnon BEC:
\begin{equation}
\omega^2(\vek{k})=c^2_{\parallel}k_z^2 +c^2_{\perp}(k_x^2+k_y^2) + M^2. 
\label{spectrum}
\end{equation}
Above and henceforth we set $\unitvec{n} = \unitvec{z}$ and $\mathbf{H} = \cos \lambda \unitvec{z} - \sin \lambda \unitvec{y}$. The small mass of the NG mode arises due to the symmetry violating RF field $H_{\rm rf} \ll H$:
\begin{equation}
 M^2 =\frac{\Omega_P^2}{8}\frac{H_{\rm rf}}{H}   (1- 5 \cos^2\lambda) \sin\beta  \,.
\label{PhononMass}
\end{equation}
This phonon mass has been measured in $^3$He-B \cite{Dmitriev2005,Skyba2008,Skyba2012} and we note that
the phonon mass is absent in pulsed NMR experiments, where the coherent precession is free.

The anisotropic ``speed of light" for phonons in magnon BEC in polar $^3$He is
\begin{equation}
c^2_{\parallel,\perp}= \frac{\epsilon''}{2}  (1- \cos\beta)(3-\cos\beta) 
K_{\parallel,\perp}\,.
\label{speeds}
\end{equation}
In terms of spin-wave velocities $c^2_{\parallel,\perp S}$ in Eq.(\ref{MagnonSpectrum}):
\begin{equation}
c^2_{\parallel,\perp}= \frac{1}{16}\frac{\Omega_P^2}{\omega_L^2}   (1- 5 \cos^2\lambda) (1- \cos\beta)(3-\cos\beta) c^2_{\parallel,\perp S}\,.
\label{speeds2}
\end{equation}
It follows that in the Ginzburg-Landau regime and weak coupling theory, also $c^2_\parallel =  3 c^2_\perp$. 

The important property of the spectrum is  that the dispersion changes sign for $\epsilon''(n, \lambda)$ at  $1- 5 \cos^2\lambda=0$. The same threshold has been calculated and observed in the disordered Larkin-Imry-Ma state of $^3$He in aerogel \cite{Dmitriev2010}.  Clearly this implies an instability of the condensate as function of the parameter $\lambda$, the angle between the magnetic field and the axis of anisotropy of the aerogel.
Here we interpret this as the transition from a Minskowski to Euclidean signature metric for the dynamical phonon modes. To see this in more detail, we compute the linear homogenous equations of motion for the phonons in the presence of counterflow in the next section. In Section 
\ref{Sec:HamiltonianToAction} ,
we complement this with a more careful analysis of the dynamical equations of motion for small deviations of the magnon BEC around equilibrium at 
$\epsilon'(n_0) =\mu - \omega_L$.

\section{Acoustic metric, ergoregion and horizon}
\label{Sec:acoustic}

\subsection{Acoustic phonon metric}

In the presence of a counterflow velocity ${\bf w}$ (say, in a rotating cryostat, or due to a spin current), a source term for the current is added to the free energy:
 \begin{align}
 F_{\rm cf}= \int d^3\vek{r}~n\nabla\alpha \cdot {\bf w} \,.
\label{cf}
\end{align}
The wave equation is modified and for constant ${\bf w}$ and $c^2_{\parallel,\perp}$, one can identify the following effective ``acoustic" metric:
%\begin{widetext}
\begin{align}
0 = \frac{\partial^2\alpha}{\partial t^2} -  
  2{\bf w} \cdot \nabla \frac{\partial\alpha}{\partial t} + ({\bf w} \cdot \nabla)^2 \alpha- \nonumber\\
 c_\parallel^2 \frac{\partial^2\alpha}{\partial z^2}
-c_\perp^2 \frac{\partial^2\alpha}{\partial x^2} -c_\perp^2 \frac{\partial^2\alpha}{\partial y^2}  
+M^2 \alpha \label{WaveEquationW}\\ 
\equiv g^{\mu\nu} \nabla_\mu \nabla_\nu \alpha + M^2 \alpha \,. \nonumber
\end{align}
%\end{widetext}
This definition of $g^{\mu\nu}$ works only for a homogenous equilibrium state since the full wave equation for a massless scalar field $\alpha$ in the background metric $g_{\mu\nu}$ is
\begin{align}
g^{\mu\nu}\hat{\nabla}_{\mu} \hat{\nabla}_{\nu} \alpha = \frac{1}{\sqrt{g}} \nabla_{\mu}(\sqrt{g}g^{\mu\nu}\nabla_{\nu}\alpha) = 0,
\end{align}
where $\hat{\nabla}$ is the covariant derivative corresponding to $g_{\mu\nu}$, see Eq. \eqref{ActionQuadratic3} below. For constant parameters in the metric the contravariant metric is
\begin{equation}
g^{00}=1\,\, , \,\, \, 
g^{0i}=-w^i \,\, , \,\, \, 
g^{ij} = w^iw^j - (c_{\parallel}^2-c_{\perp}^2)\hat{z}^i\hat{z}^j - c_{\perp}^2\delta^{ij}.
\label{metric}
\end{equation}
This form of the metric corresponds to the Hamiltonian or ADM formalism of general relativity \cite{ADM} with the shift vector $N^i = w^i$ and the gauge fixed lapse function $N=1$. This fixed-gauge  metric is natural for condensed matter analogies of general relativity, since the spectrum is obtained in the laboratory frame of the condensate. In this gauge $g^{00}=1/N^2=1$, and the metric determinant is:
\begin{equation}
g= -\frac{1}{c_\perp^4 c_\parallel^2}    \,.
\label{MetricDeterminant}
\end{equation}
This also confirms the correct choice of the gauge, since in the laboratory frame the ``speed of light" is anisotropic, and thus there is no unique definition of the propagation speed of the Goldstone modes. However, such metric is not suitable when the determinant changes sign.

Two surfaces related to the metric \eqref{metric} are of interest: the surface at which $c_\perp^2({\bf r})=w^2({\bf r})$ where $g_{00}$ and $g^{ww}$  cross  zero;
 and the surface where  $c_\perp^2({\bf r})$ and $c_\parallel^2({\bf r})$ cross zero and become negative. Let us start with the first one. The surface at which $c_\perp^2({\bf r})-w^2({\bf r})$ crosses zero is either a horizon or an ergosurface, depending on the  orientation of the interface with respect to flow velocity ${\bf w}$. The ergosurface takes place for the circular flow 
$ {\bf w}=v({\rho})\hat{\bf\phi}$. Assuming that $\lambda$ is also axisymmetric, one has for the covariant acoustic metric $g_{\mu\nu}$:
\begin{align}
ds^2 &=g_{\mu\nu} dx^\mu dx^\nu 
\nonumber
\\
&= dt^2- \frac{\rho^2}{c_\perp^2({\rho})} \left( d\phi -  \frac{v({\rho})}{\rho}dt \right)^2-\frac{d\rho^2}{c_\perp^2({\rho})} -\frac{dz^2}{c_\parallel^2({\rho})}\,.
\label{covariant}
\end{align}
The ergosurface is at $c_\perp^2({\rho})= v^2({\rho})$, where 
$g_{00}$ and
$g^{\phi\phi}$  cross zero. 

Note that for such metric there are no closed time-like curves. For the closed time-like curve to  exist it is necessary to have $g_{\phi\phi}>0$. This occurs in the G\"odel Universe, where the corresponding acoustic metric has been discussed \cite{Kajari2004}. At such surface  $g_{\phi\phi}$ and $g^{00}$ cross zero, and thus behind such surface the closed time-like curves appear. The vacuum in this region is unstable, as can be seen from the spectrum of photons.
Similar instability takes place in our case when $c_\perp^2({\bf r})$ and $c_\parallel^2({\bf r})$ in Eqs.(\ref{speeds}) and (\ref{speeds2})  become negative.

\subsection{Minkowski-to-Euclidean signature change of the effective Nambu-Goldstone metric.}
\label{Sec:Euclidean}

 Let us consider the surface, where $c_\perp^2({\bf r})=0$ and $c_\parallel^2({\bf r})=0$. At this surface $g$ in Eq.(\ref{MetricDeterminant}) crosses infinity and changes sign, i.e. the Minkowski signature transforms to the Euclidean one. Such transformation via 
$g=\infty$ and also via $g=0$ has been discussed for the metric induced on a cosmic string in the presence of black hole, where it was mentioned that  such spacelike space-time does not represent any solution for a physical cosmic string \cite{Frolov2004}. 
In our case, when $\epsilon''(n,\lambda) <0$, the Euclidean space-time for the Goldstone mode signals the instability of the coherent precession. Such instability has been discussed and has been observed in the A-phase, see the review paper \cite{Fomin1990}. In the polar phase the instability, $c^2_{\parallel,\perp}<0$, takes place when
\begin{equation}
 \tan\lambda <  2 \,.
\label{threshold}
\end{equation}
At $\tan\lambda_c = 2$, there is a transition from Minkowski signature at $\tan\lambda> 2$ to Euclidean signature at $\tan\lambda< 2$. 
Such change in the signature of the phonon metric   has been discussed for the Bose gas at the transition between the repulsive and attractive interaction of bosons \cite{Visser2007}.
However, in our case the (small) mass becomes simultaneously tachyonic, $M^2<0$.
Excitations with such a spectrum can be called tachyonic ghosts \cite{Arkani2004}.
The instability is much stronger than in the ergoregion, but experimentally the lifetime of the unstable vacuum can be made long enough near the threshold of instability.

When the effective metric depends on coordinates, its behavior in the instability region differs from what follows from the linear equations. This can be seen from the consideration of the full action of quadratic deviations from equilibrium, discussed in the next section.

\section{Acoustic metric of quadratic deviations and signature change}
\label{Sec:HamiltonianToAction}

We now wish to calculate the signature change of the full acoustic metric for the phonons by considering the quadratic action of deviations around equilibrium.

The phonon Hamiltonian \eqref{eq:Hamiltonian0} with the counterflow term is
%\begin{widetext}
\begin{align}
H(\alpha,n)&=\int d^3 \vek{r} \bigg(  \epsilon(n) + (\omega_L-\mu) n \label{Hamiltonian}\\
&\phantom{=}+ \frac{1}{2} \gamma^{ij}(n) \nabla_i \alpha  \nabla_j \alpha + n{\bf w}\cdot \nabla \alpha + f_{\rm sb}(n,\alpha)\bigg)\,,
\nonumber
\end{align}
%\end{widetext}
the action follows as
\begin{align}
S=\int dt  \left(\int d^3 \vek{r} \, n \dot \alpha -H(\alpha,n)\right)\,
\label{Action}
\end{align}
and defines the quantum mechanical path-integral kernel $Z_{i\to f}= \langle f \vert T[e^{-\im \int_0^T H(t)}] \vert i \rangle = \int_{\alpha_0=i}^{\alpha_T=f} \mathcal{D}\alpha e^{\im S}$ with saddle point solutions corresponding to the classical equations of motion. 

To obtain the action for the phonon modes, we need the quadratic form of deviations $n=n_0+\delta n$ and $\alpha= \alpha_0+\delta \alpha$ around some equilibrium state $n_0(\vek{r})$, where $\frac{\delta H}{\delta n}\vert_{n=n_0}=0$ and $\frac{\delta H}{\delta \alpha}\vert_{\alpha=\alpha_0}=0$. Ignoring the small mass term $f_{\rm sb}(n,\alpha)$ and the equilibrium spin current $ \nabla\alpha_0$, i.e. expanding around constant $\alpha_0$ and $\epsilon'(n_0)=\mu-\omega_L$, leads to the canonical relation
\begin{equation}
\delta n = \frac{1}{\epsilon''(n_0)}  (\delta \dot \alpha - {\bf w}\cdot \nabla \delta\alpha) \,.
\label{deltan}
\end{equation}
Then the action for the quadratic deviations $\delta \alpha\equiv \alpha$ is
\begin{align}
S &=  \frac{1}{2}  \int dt \int d^3\vek{r}  \left(  \frac{1}{\epsilon''}  (\dot \alpha - {\bf w}\cdot \nabla \alpha)^2
-\gamma^{ij} \nabla_i \alpha  \nabla_j \alpha \right).
\label{ActionQuadratic1}
\end{align}

\subsection{Stable Minkowski region $\epsilon''(n_0,\lambda)>0$}

In the convex region, where $\epsilon''(n_0,\lambda)>0$, i.e. $\tan \lambda >2$, the spin-orbit interaction reproduces the  repulsive interaction of magnons, and the magnon BEC is stable. The quadratic action can be written in terms of effective metric $g ^{\mu\nu}$ for the scalar field $\alpha$:
\begin{align}
S&\equiv
 \frac{1}{2} \int dt \int d^3\vek{r}  \, \tilde g^{\mu\nu} \nabla_\mu \alpha  \nabla_\nu \alpha 
    \label{ActionQuadratic2}
 \\
&\equiv
 \frac{1}{2} \int dt \int d^3\vek{r}  \, \sqrt{-g} g ^{\mu\nu} \nabla_\mu \alpha  \nabla_\nu \alpha 
\,.
\label{ActionQuadratic3}
\end{align}
Here the matrix $\tilde g^{\mu\nu}$ follows directly from \eqref{ActionQuadratic1} along with the inverse $\tilde{g}_{\mu\nu}$,
\begin{align}
\tilde g^{00} = \frac{1}{\epsilon''}, \quad 
\tilde g^{0i} =- \frac{w^i}{\epsilon''},  \quad
\tilde g^{ij} =- \gamma^{ij} + \frac{w^i w^j}{\epsilon''}, \\%\quad  
%\tilde g =   \frac{\epsilon''}{\gamma}    \,. \\
\label{MetricTilde}
\tilde{g}_{00} = \epsilon'' - \gamma_{ij}w^i w^j,\quad \tilde{g}_{0i} = - \gamma_{ij}w^j,\quad \tilde{g}_{ij} = -\gamma_{ij}. 
\end{align}
where $ \tilde{g} \equiv \det \tilde{g}_{\mu\nu} = -\frac{\epsilon''}{\gamma}$ and ${\gamma}$ is the determinant of the matrix $\gamma^{ij}$. Comparing Eqs.(\ref{ActionQuadratic2}) and (\ref{ActionQuadratic3}) one obtains the effective metric:
\begin{equation}
 g^{\mu\nu}=\sqrt{-\tilde g} \tilde g^{\mu\nu}  =   \left(\frac{\epsilon''}{\gamma}\right)^{1/2}   \tilde g^{\mu\nu}\,\, , \,\, \, 
 \sqrt{-g} = \frac{1}{  \sqrt{-\tilde g} } 
 \,.
\label{EffectiveMetric1}
\end{equation}
or
\begin{equation}
  g^{00}=\frac{1} {\sqrt{\gamma\epsilon''} }
  \,\, , \,\, \, 
   g^{0i}= -\frac{w^ i} {\sqrt{\gamma\epsilon''} }
     \,\, , \,\, \, 
     g^{ij}= \frac{w^ i w^j}{\sqrt{\gamma\epsilon''}}- \gamma^{ij}
 \sqrt{\frac{\epsilon''}{\gamma} }    \,.
\label{EffectiveMetric2}
\end{equation}

\subsection{Unstable Euclidean region $\epsilon''(n_0, \lambda)<0$}

In the region $\tan \lambda <2$, $\epsilon''<0$ and the spin-orbit interaction is concave and the magnon BEC becomes unstable. The above description for the acoustic metric is not valid, since effective metric $g^{\mu\nu}$  Eq.(\ref{EffectiveMetric2}) becomes imaginary, while
the motion equations are still real. Moreover, the determinant of the metric changes sign and the metric signature becomes Euclidean.
From Eq. \eqref{ActionQuadratic1} the correct form of the action follows as
%\begin{widetext}
\begin{align}
S &\equiv
 -\frac{1}{2} \int dt \int d^3\vek{r}  \, \tilde g_E^{\mu\nu} \nabla_\mu \alpha  \nabla_\nu \alpha \nonumber\\
 &\equiv
- \frac{1}{2} \int dt \int d^3\vek{r}  \, \sqrt{g_{\rm E}} g_{\rm E}^{\mu\nu} \nabla_\mu \alpha  \nabla_\nu \alpha 
\,.
\label{ActionEucl}
\end{align}
%\end{widetext}
Comparing Eqs. \eqref{ActionQuadratic1} and \eqref{ActionEucl} one obtains the effective metric
\begin{equation}
 g_{\rm E}^{\mu\nu}=\sqrt{\tilde g} \tilde g^{\mu\nu}  =   \left(-\frac{\epsilon''}{\gamma}\right)^{1/2}   \tilde g^{\mu\nu}\,\, , \,\, \, 
 \sqrt{g_{\rm E}} = \frac{1}{  \sqrt{\tilde g} } 
 \,,
\label{EffectiveMetric1Euc}
\end{equation}
which now has Euclidean signature,
\begin{align}
  g_{\rm E}^{00}=\frac{1} {\sqrt{-\gamma\epsilon''} }
  \,\, , \,\, \, 
   g_{\rm E}^{0i}= -\frac{w^ i} {\sqrt{-\gamma\epsilon''} }, \nonumber\\  
     g^{ij}_{\rm E}= \frac{w^ i w^j}{\sqrt{-\gamma\epsilon''}}+ \gamma^{ij}
 \sqrt{-\frac{\epsilon''}{\gamma} }    \,.
\label{EffectiveMetric2Euc}
\end{align}

The Euclidean signature corresponds to complex phonon frequencies $\omega^2(\vek{k})<0$ of the condensate and therefore makes the condensate unstable to the phonon modes. We stress that the Euclidean metric $g_{{\rm E},\mu\nu}$ does not correspond to the imaginary time thermal partition function of the condensate but instead to the the dynamical phonon modes around the equilibrium state inherited from the stable equilibrium state for $\epsilon''(n_0,\lambda)>0$. Dynamics is governed by the Euclidean signature metric in the non-equilibrium region where the expansion around the unstable condensate is still valid.

When the instability to phonon NG modes develops at $\tan \lambda <2$, it is cut-off at higher energies by the quartic term in Eq. \eqref{WaveEquation2} that have the usual, Minkowski signature in the dispersion. Similar role of the higher-than-quadratic terms has been discussed in Refs. \cite{Visser2007,Arkani2004}. The magnon BEC in the unstable region can be considered as an effective non-relativistic version of a ``ghost condensate" similar to that of Ref. \onlinecite{Arkani2004}.

\section{Outlook}
\label{sec:Outlook}

Superfluid $^3$He has stable region of  a magnon BEC when the precession averaged spin-orbit interaction $f_{\rm so} = \epsilon(n)$ is convex, which plays the role of repulsive magnon-magnon interaction. This is %an out-of equilibrium 
the long-lived state of coherent precession formed after optical magnons are pumped with a RF pulse.  The magnon number $N=({\cal S}-{\cal S}_z)/\hbar$  determines the global frequency of precession, which plays the role of chemical potential, $\omega\equiv \mu_{\rm BEC}$, while in continuous RF magnetic field the magnon BEC is stabilized with $\mu_{\rm BEC}=\omega_{\rm rf}$.  The precession phase $\alpha$ corresponds to the conjugate degrees of freedom of the condensate. The most well-known example of this is the HPD in $^3$He-B. Here we have discussed the HPD magnon BEC in the polar phase of $^3$He. 

At the critical value $\tan \lambda_c = 2$ of the angle $\lambda$ between the axis of anisotropy $\unitvec{n}$ of the aerogel and the static magnetic field $\vek{H}$, there is a transition from repulsive to attractive magnon spin-orbit interaction.  When $\tan\lambda < 2$, we have $\epsilon''(n_0,\lambda)<0$ around equilibrium and the magnon BEC becomes unstable. The magnetic field angle $\lambda$ can be tuned in experiments continuously from the stable to the unstable region. Such transition has been discussed for $^3$He-A \cite{Gurgenishvili1985,BunkovVolovik1993} and observed in the disordered Larkin-Imry-Ma state of $^3$He in aerogel \cite{Dmitriev2010}.  Here we  considered this transition and condensate instability in terms of  the effective phonon  metric.

In magnon superfluids there are two effective metrics. One is the metric for the propagaing magnons in Eq.(\ref{EffectiveMagnonMetric}). The other one is the effective metric experienced by phonons propagating in the magnon BEC in Eqs.(\ref{metric}) and (\ref{EffectiveMetric2}). This is an analog of acoustic metric introduced by Unruh \cite{Unruh1981}.
The  transition between the repulsive and attractive spin-orbit interaction corresponds to the transition between Minkowski and Euclidean signature of the phonon  metric. The Euclidean metric 
$g_{{\rm E},\mu\nu}$ in Eq.(\ref{EffectiveMetric2Euc}) does not correspond to the imaginary time \emph{thermal} partition function, but is relevant  for the \emph{dynamic} phonon modes of the condensate.

 The signature change of the metric takes place in many models of the early universe in cosmology, for quantum gravity and for cosmic strings \cite{Witten1981,Uzan2013,Frolov2004}. 
The transition of the Lorentzian signature to Euclidean triggers a ghost instability of the quantum vacuum \cite{Arkani2004, Motohashi2015}.
This corresponds to instability of  the magnon BEC  in the Euclidean region, where the BEC decays as a false vacuum.
Depending on the experimental conditions, the decay rate may be long enough to simulate different mechanisms of the decay of the false magnon vacuum.

We also found a difference in the phenomenology of magnon superfluidity for the isotropic $^3$He-B and the polar phase with an easy axis anisotropy (see the Appendix). In $^3$He-B the phenomenology corresponds to a BEC of magnons with the inertial mass $M =  \hbar\omega_L/2$, which coincides with the  invariant mass (gap) of the optical magnon. In the polar phase the inertial mass of the effective bosons is twice as large, $M_{\rm eff} =  \hbar\omega_L$. However, the effective metric for these bosons coincides with the metric of the optical magnons.

{\bf Acknowledgements}. This work has been supported by the European Research Council (ERC) under the European Union's Horizon 2020 research and innovation programme (Grant Agreement No. 694248).

%\appendix*
\section*{Appendix. Effective bosons in the precessing magnon BEC} \label{EffectiveMagnons}
Here  we discuss the phonon Hamiltonian $H(\alpha,n) = H_{\rm BEC} - \mu N = F_{\rm BEC}$ of the magnon condensate in the polar and B-phase of $^3$He. 
The effective condensate free energy $F_{\rm BEC}$ is the precession averaged free energy of the superfluid in the London limit.  In the $n \ll n_{\max}$ limit, it has the conventional form of the Ginzburg-Landau (GL) free energy of the BEC:
\begin{align}
F_{\rm BEC} =  \int d^3 \vek{r}~ \frac{1}{2}\frac{n}{S}K_{ij} \nabla_i \alpha \nabla_j \alpha 
+ \frac{1}{2 S} \frac{K_{ij}}{4n}\nabla_i n \nabla_j n \\
+ (\omega_L-\mu)n + \epsilon(n) \nonumber\\
= \int d^3\vek{r}~\frac{1}{2m_{\parallel}} \vert \nabla_{\parallel} \Psi \vert^2 
+ \frac{1}{2m_{\perp} }\vert \nabla_{\perp}  \Psi\vert^2+(\omega_L-\mu)\vert \Psi\vert^2 \label{GLgradient}\\ 
+\epsilon(\vert \Psi\vert^2),  \Psi =\sqrt{n} e^{i\alpha}, \vert\Psi\vert^2= n=S(1-\cos\beta)\,. \nonumber
\end{align}
Here $m_{\parallel}$ and $m_{\perp}$ are effective inertial masses with
\begin{align}
 (m^{-1})_{ij} = \frac{K_{ij}}{S} = \frac{g^{ij}_S} {\hbar\omega_L}\,.
\label{BosonMass}
\end{align} 
The gradient term in $n$ corresponds to the vacuum pressure and gives the fourth-order correction to the spectrum.  As distinct from the conventional Bose gas, in our case the masses in the GL free energy do not necessary coincide with the true magnon masses. They can be considered as effective masses of the bosons forming the magnon BEC.
The inertial mass in Eq.(\ref{BosonMass}) corresponds to the effective invariant mass $M_{\rm eff}=\hbar\omega_L$ of the boson, which 
 coincides with the invariant mass of the magnon in Eq.(\ref{OpticalMagnonMass}).

Let us compare this with the magnon BEC of $^3$He-B in Ref. \cite{BunkovVolovik2013}. For small $n \ll n_{\rm max}$, i.e. for $\beta\ll 1$, the gradient terms of $\alpha$ in the free energy correspond to the kinetic energy of the magnon BEC: 
 \begin{equation}
\langle f^{\rm B}_{\rm grad}(n\to 0) \rangle= \frac{1}{2}\rho_{{\rm s}ij}v_{{\rm s}i} v_{{\rm s}j} \,,
\label{HydrodynamicGradEnergy}
\end{equation}
where  $\rho_{{\rm s}ij}$ is the tensor of anisotropic superfluid density and  $v_{{\rm s}i} $ is the superfluid velocity of magnon superfluid: 
\begin{equation}
\rho_{{\rm s}ij} =nm_{\textrm{B}, ij} ~~,~~ v_{{\rm s}i}  =\hbar \left(m_{\rm B}^{-1}\right)_{ij}  \nabla_j \alpha \,,
\label{SuperfluidDensity}
\end{equation} 
 where  the matrix of masses $m_{\textrm{B},ij}$ for $\beta \ll 1$ is
\begin{eqnarray}
 (m_{\rm B}^{-1})_{ij} = \frac{2g^{ij}_S} {\hbar\omega_L}
  \,.
\label{HeB_superfluid_mass}
\end{eqnarray}
In this case the inertial mass of effective boson corresponds to the invariant magnon mass % (gap) 
$M_+ =  \hbar\omega_L/2$ of the spectrum in the B-phase, see Eq.(11) in the supplementary material to Ref.\onlinecite{Autti2016b}
The mass-supercurrent  expressed via $\alpha$ is
\begin{equation}
J_i= \frac{\delta F^{\rm B}}{\delta v_{{\rm s}i} }=\hbar n \nabla_i\alpha \,,
\label{MassCurrent2}
\end{equation}   
and similarly coincides with the linear momentum of the magnon condensate. 
It follows that the out-of-equilibrium magnon BEC in $^3$He-B at small $n$ is very similar to the conventional BEC of bosonic particles with the inertial mass equal to the invariant mass of the optical magnon $M_+=  \hbar\omega_L/2$.  In contrast, in the polar phase the effective boson (the optical magnon) has the twice larger mass, $M_{\rm eff}=\hbar\omega_L=M_+$. 

This can be directly verified from $F_{\rm BEC}$ at finite momenta, when the phonon NG approximation is not valid. Consider the spectrum of phonons in the intermediate
regime $1/\xi_{\rm BEC}  \ll   k \ll mc_S$, where $\xi_{\rm BEC}\sim c_S/\Omega_P$. In the BEC theory this spectrum should approach the spectrum of the bosons forming the BEC. In this imit the quartic term in Eq.(\ref{WaveEquation2})   is dominating, and for
small $\tilde n\ll 1$ it gives:
\begin{equation}
  \omega_{\rm phonon} \rightarrow  \frac{K_{\parallel,\perp}k_{\parallel,\perp}^2}{2S} = \frac{c^2_{\parallel,\perp S}k_{\parallel,\perp}^2}{2\omega_L}  
  = \frac{k_{\parallel,\perp}^2}{2m_{\parallel,\perp}} \,,
\label{Largek}
\end{equation}
where again $c_{\parallel,\perp S} = \gamma_m^2 K_{\parallel,\perp}/\chi$. This does match the spectrum of optical magnons at $\Omega_P  \ll c_S k \ll \omega_L$:
\begin{equation}
  \omega^{\rm optical}_{\parallel,\perp} -\mu \rightarrow \frac{c^2_{\parallel,\perp S}k_{\parallel,\perp}^2}{2\omega_L}   \,,
\label{optical}
\end{equation}
where $\mu =\omega_L$.

\end{document}